\documentclass[12pt]{article}
\usepackage{times}
\usepackage{geometry}
\usepackage{subfigure}
\usepackage{enumitem,amssymb}
\usepackage{natbib}
\usepackage{ragged2e}
\usepackage{color}
\newlist{thematic}{itemize}{8}
\setlist[thematic]{label=$\square$}
\usepackage{pifont}
\usepackage{graphicx}
\newcommand{\cmark}{\ding{51}}%
\newcommand{\done}{\rlap{$\square$}{\raisebox{2pt}{\large\hspace{1pt}\cmark}}%
\hspace{-2.5pt}}

\newcommand{\JWST}{\textit{JWST}}

\begin{document}
\raggedright
\huge
Astro2020 Science White Paper \linebreak

EUV influences on exoplanet atmospheric stability and evolution \linebreak
\normalsize

\noindent \textbf{Thematic Areas:} \hspace*{60pt} \done~Planetary Systems \hspace*{10pt} $\square$ Star and Planet Formation \hspace*{20pt}\linebreak
$\square$ Formation and Evolution of Compact Objects \hspace*{31pt} $\square$ Cosmology and Fundamental Physics \linebreak
  $\square$  Stars and Stellar Evolution \hspace*{1pt} $\square$ Resolved Stellar Populations and their Environments \hspace*{40pt} \linebreak
  $\square$    Galaxy Evolution   \hspace*{45pt} $\square$             Multi-Messenger Astronomy and Astrophysics \hspace*{65pt} \linebreak
  
\textbf{Principal Author:}

Name: Allison Youngblood
 \linebreak						
Institution:  NASA Goddard Space Flight Center
 \linebreak
Email: allison.a.youngblood\@nasa.gov
 \linebreak
Phone:  (301) 614-5729
 \linebreak
 
\textbf{Co-authors:} Kevin France (University of Colorado), Tommi Koskinen (University of Arizona), Luca Fossati (IWF, Graz), Ute Amerstorfer (IWF, Graz), Herbert Lichtenegger (IWF, Graz), Jeremy Drake (Harvard SAO), James Mason (NASA GSFC), Brian Fleming (University of Colorado), Joel Allred (NASA GSFC), Zachory Berta-Thompson (University of Colorado), Vincent Bourrier (University of Geneva), Cynthia Froning (University of Texas), Cecilia Garraffo (Harvard CfA), Guillaume Gronoff (NASA LRC), Meng Jin (Lockheed Martin), Adam Kowalski (University of Colorado), Rachel Osten (STScI, JHU)
\linebreak

\textbf{Abstract:} The planetary effective surface temperature alone is insufficient to characterize exoplanet atmospheres and their stability or evolution. Considering the star-planet system as a whole is necessary, and a critical component of the system is the photoionizing stellar extreme ultraviolet emission (EUV; 100-912 \AA). EUV photons drive atmospheric mass loss through thermal and nonthermal processes, and an accurate accounting of the EUV energy deposition in a planet's energy budget is essential, especially for terrestrial habitable zone planets and close-in gaseous planets. Direct EUV observations of exoplanet host stars would require a new, dedicated observatory. Archival observations from the \textit{EUVE} satellite, models, and theory alone are insufficient to accurately characterize EUV spectra of the majority of exoplanet host stars, especially for low-mass stars.

\pagebreak

\section*{EUV Photons Drive Mass Loss}
\begin{figure}
    \centering
    \subfigure{
    \includegraphics[width=\textwidth]{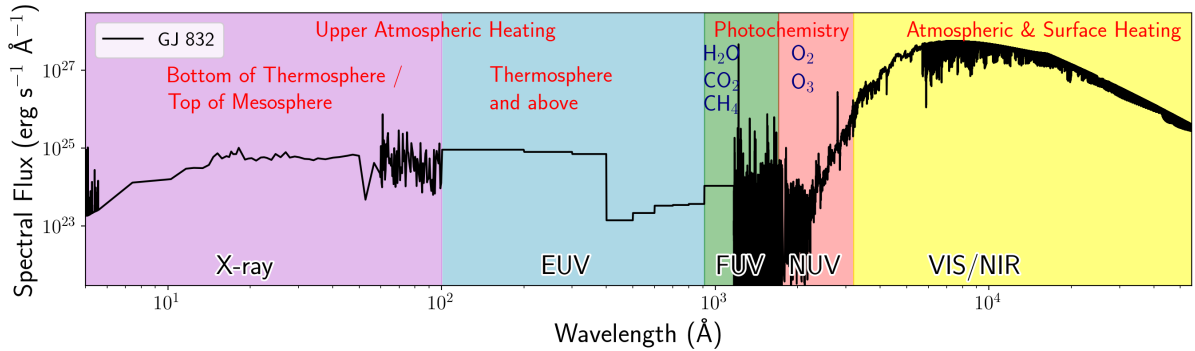} }
    \caption{\textbf{Different spectral regions have different effects on planet atmospheres.} The spectral energy distribution for the M dwarf GJ 832 (d=4.97 pc) from the MUSCLES Treasury Survey \citep{France2016,Youngblood2016b,Loyd2016}. MUSCLES, a widely-used database of M and K dwarf SEDs, used the empirical scaling relations of Linsky et al. (2014) to estimate the EUV spectrum from the ISM-corrected Lyman-$\alpha$ emission. }
    \label{fig:GJ832_SED}
\end{figure}

Different parts of a star's spectral energy distribution (SED) drive heating and chemistry in different layers of a planet's atmosphere due to the wavelength dependence of atomic and molecular photoabsorption cross sections (Figure~\ref{fig:GJ832_SED}). Optical and near-infrared photons heat the surface and troposphere. NUV, FUV and X-ray photons are absorbed higher in the atmosphere where they photodissociate molecules; the resulting lighter atoms and molecules are more susceptible to escape from the planet via thermal and non-thermal processes.  EUV photons are absorbed even higher in the atmosphere where they ionize atoms and molecules.  The resulting free electrons further contribute to thermal and non-thermal atmospheric escape processes.  The stability of Earth-like atmospheres depends critically on X-ray and UV radiation, with EUV radiation playing a critical role at the highest altitude in the atmosphere. 
\medskip

M dwarf exoplanets are particularly prone to EUV-driven escape given their long pre-main-sequence phase characterized by EUV luminosity factors of 10 - 100 times higher than the main sequence value (e.g., \citealt{Ribas2016}). This enhanced EUV phase on M dwarfs can last up to two billion years longer than that of Sun-like stars \citep{Ramirez2014}, well into or beyond the era in which life emerged on Earth.

\section*{Mechanisms of EUV-driven atmospheric escape}
EUV photons ionize atoms and molecules, creating a population of photoelectrons. Electrons collisionally heat the surrounding gas, increasing the scale height of the atmosphere and possibly leading to the onset of a hydrodynamic outflow (i.e., hydrodynamic escape), or even enhancing the upper atmosphere's susceptibility to pick-up ion loss from the stellar wind, even in the presence of an intrinsic planetary magnetic field. If a hydrodynamic outflow is sufficiently rapid, heavier elements like O, C, and Si can also be dragged along through collisions with the lighter hydrogen \citep{Vidal-Madjar2004,Linsky2010,Koskinen2014,Ballester2015,Ben-Jaffel2013,Dong2018}. In gas giants, elements as heavy as Mg and Fe may also ionize and hydrodynamically escape under the action of stellar X-ray and EUV heating (e.g., \citealt{Fossati2010,Haswell2012}). 
\medskip

\cite{Amerstorfer2017} showed that 20$\times$~greater solar EUV flux, which would have occurred when the Sun was 300-900 Myr old \citep{Tu2015}, can lead to 10$\times$~greater O loss rates and 90$\times$ greater C loss rates on Mars by increasing the suprathermal or ``hot" population of these atoms. Photodissociation of CO is a major contributor to the hot atom population at high EUV fluxes, but other sources like dissociative recombination of O$_2^+$ from photoelectrons are significant at lower EUV fluxes.
\medskip

Free electrons created by ionizing EUV photons can attain altitudes much greater than ions, producing an ambipolar electric field which leads to an ionospheric outflow. Large amounts of oxygen and nitrogen, important atoms for habitability and biogenesis, can therefore be lost as O$^+$ and N$^+$ under strong enough irradiation conditions \citep{Kulikov2006,Lichtenegger2016,Airapetian2017,Dong2017,Dong2017a,Garcia-Sage2017}. EUV spectra are necessary to quantify the deposition of escape-driving energy into exoplanet atmospheres, particularly because the photoabsorption cross sections are highly wavelength dependent. 

\section*{The importance of \textit{spectrally resolved} EUV observations for exoplanet host stars}
Spectrally resolved EUV observations compared to broadband and narrowband photometry are particularly beneficial because the EUV spectra of cool stars are dominated by a forest of narrow emission lines with relatively weak continua.  Due to the dependency of typical absorption cross sections on photon energy, radiation in different emission lines penetrates to different altitudes in the atmosphere, modifying the energy balance, ionization rates, and photochemistry at these altitudes in ways that cannot be captured by models based on single wavelength or broadband irradiation approximations.  In a nutshell, the deeper the radiation penetrates, the less likely it is to contribute significantly to mass loss, while still having potential to alter photochemistry.  EUV radiation is absorbed higher in the atmosphere than UV or X-rays. For both photochemistry and escape, it is therefore important to include the EUV spectrum of the host star in models of planetary atmospheres instead of a single-valued flux. Recent work in the solar system provides interesting context.  For example, \cite{Kim2014} show that a combination of a high-resolution solar spectrum and electronic band absorption cross section of H$_2$ leads to the penetration of solar EUV radiation much deeper into Saturn's atmosphere, past the thermosphere, producing ionization of methane and the production of complex hydrocarbons at rates far exceeding previous estimates based on broadband spectra and cross sections.  Similarly, \cite{Lavvas2011} found that using high resolution N$_2$ band cross sections with the high-resolution \textit{SOHO}/SUMER spectrum of the Sun led to deeper penetration of photons into Titan'€™s upper atmosphere.

\section*{Archival EUV observations, models, and theory are insufficient}

EUV spectra of exoplanet host stars are scarce, particularly for M dwarfs. The only previous 
EUV astronomy mission, \textit{EUVE} \citep{Bowyer1991}, obtained spectra of 10 cool mainsequence stars, including 5 M dwarfs, all of which are active flaring stars and not representative of
the majority of the exoplanet host stars we know of today \citep{France2013}. No EUV spectra
exist of the relatively inactive M dwarfs that will be optimal for biosignature searches with
\JWST~and ELTs, putting astronomers at a disadvantage for understanding how potential habitats
evolve with time and designing truly informative biosignature searches.
\medskip

\begin{figure}
    \centering
    \includegraphics{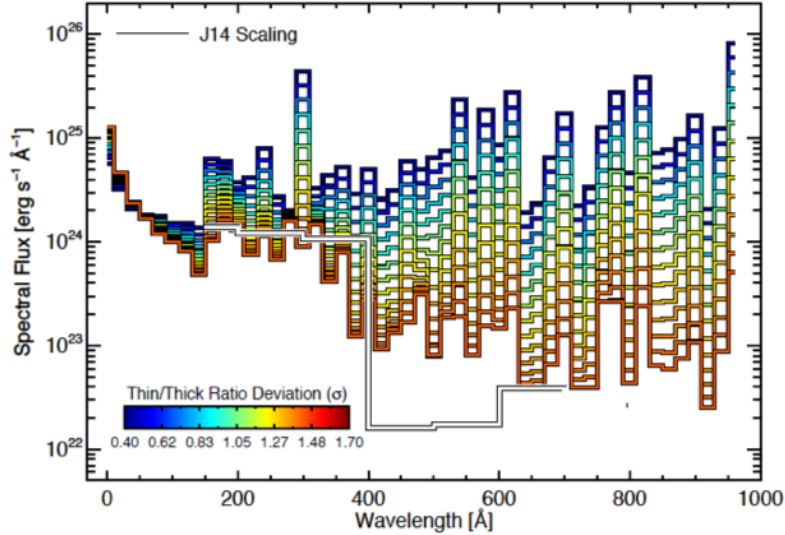}
    \caption{\textbf{DEMs constrained only from X-rays cannot precisely constrain EUV spectra at $\lambda$~$\textgreater$ 150 \AA.} Synthetic EUV spectra computed from plausible model DEMs for the fully convective M dwarf LHS 248 (M6.5 V) based on constraints from \textit{Chandra} HRC-S observations. The models are color-coded according to their goodness of fit to the \textit{Chandra} data. The flux predicted by the Linsky et al. (2014) empirical scaling based on the Lyman-$\alpha$ flux of Proxima Centauri, scaled to the radius of LHS 248 is also shown (denoted J14). While the model flux is well constrained for wavelengths $\textless$ 150 \AA, the uncertainties in the predicted flux given by the spread in the models grows to more than an order of magnitude by 400 \AA. From Drake et al., ApJ, submitted.}
    \label{fig:EUV_models_comparison}
\end{figure}

Stellar models that can predict the EUV emission for individual stars by including a chromosphere, transition region, and corona are in their infancy \citep{Fontenla2016,Peacock2019}. Driven by the lack of detailed models and direct EUV spectra of planet-hosting stars, the exoplanet community has developed numerous EUV scaling relations with chromospheric, transition region, and coronal emission lines available in the FUV and/or X-ray spectral regions \citep{Sanz-Forcada2011,Fontenla2011,Linsky2014,Chadney2015,Fossati2015,Louden2017,King2018,France2018}. These approaches all have significant shortcomings: while X-rays are generated in the stellar corona and FUV Lyman-$\alpha$ photons are generated in the chromosphere and transition region, solar atmosphere modeling demonstrates that the EUV photons are generated from a range encompassing chromospheric, transition region, and coronal temperatures \citep{Fontenla2014}.  Empirical EUV scaling relations and data-driven emission measure distributions can vary by a 1-3 orders of magnitude (Figure~\ref{fig:EUV_models_comparison}). The differential emission measure method, which relies on the availability of spectra covering a wide range of emission lines originating throughout every temperature of the stellar upper atmosphere, will be incomplete for the EUV spectral region even with deep X-ray and FUV spectroscopy, because plasma temperatures in the range 10$^{5.5}$-10$^{6.2}$ K are only probed in the EUV spectral region. See the white paper ``EUV observations of cool dwarf stars" led by A. Youngblood for more details about the insufficiencies of the DEM method for treating the EUV.

\section*{Time variability and the influence of flares}
Solar EUV emission varies by factors up to $\sim$10$^2$ on minute timescales due to flares and by factors of $\sim$1.3-6 on 11 year timescales due to the solar cycle \citep{Woods2012}. Different emission lines and continuua vary by different factors on all timescales. High-energy emission from M dwarfs is also highly variable, with many M dwarfs displaying UV and X-ray flare rates and energies orders of magnitude higher than solar-type stars \citep{Loyd2018}. The effect of large and frequent flares on Earth-like atmospheres is not well explored, and most studies rely on rough estimates of flares spectral responses because of the lack of direct data. For example, an estimate of the EUV response of the Great AD Leo Flare of 1985 observed with \textit{IUE} \citep{Hawley1991} indicates that the atmospheres of planets orbiting very active stars with frequent flares are never at steady state \citep{Venot2016} because the timescales for atmospheric recovery are much longer than the time between successive flares. The lack of direct constraints on stellar EUV flare frequencies and energies for stars of differing mass and activity level greatly hinders our ability to place exoplanet atmospheric evolution studies on firm ground. The required monitoring observations cannot be collected from the ground because of the lack of optical features whose behavior is intimately linked to the EUV emission.
\medskip

\section*{Conclusions}
A dedicated EUV observatory is necessary to directly measure the stellar photons that drive the evolution of exoplanetary atmospheres and determine their potential to support Earth-like conditions. We must measure the absolute EUV irradiance incident on orbiting exoplanets and quantify the contribution of stellar flares to the overall stellar EUV energy budget. With direct EUV spectroscopy, we will be able to determine if atmospheric loss rates driven by EUV photons are sufficient to alter exoplanet habitability and demographics such as the 1.5-2 R$_{\oplus}$ radius gap observed by \textit{Kepler} \citep{Fulton2017}.

\pagebreak
\bibliographystyle{proposal}
\bibliography{main.bbl}

\end{document}